\documentclass[aps, prd, nofootinbib, preprintnumbers, showpacs]{revtex4}
\usepackage{graphicx}
\usepackage{amsmath}
\usepackage{multirow}
\usepackage{bm}
\usepackage{color}
\def\fsl#1{\setbox0=\hbox{$#1$}                 
   \dimen0=\wd0                                 
   \setbox1=\hbox{/} \dimen1=\wd1               
   \ifdim\dimen0>\dimen1                        
      \rlap{\hbox to \dimen0{\hfil/\hfil}}      
      #1                                        
   \else                                        
      \rlap{\hbox to \dimen1{\hfil$#1$\hfil}}   
      /                                         
   \fi}                                         %
\newcommand{\tr}{\mbox{tr}}

\begin{document}
\title{Dynamical electroweak symmetry breaking with
superheavy quarks \\ and $2+1$ composite Higgs model}
\author{Michio Hashimoto}
  \email{michioh@post.kek.jp}
  \affiliation{Theory Center, Institute of Particle and Nuclear Studies,\\
   High Energy Accelerator Research Organization (KEK), \\
   1-1 Oho, Tsukuba, Ibaraki 305-0801, Japan}
\author{V.A. Miransky}
  \email{vmiransk@uwo.ca}
\altaffiliation[On leave from ]{Bogolyubov Institute for Theoretical 
Physics, 03680, Kiev, Ukraine.}
\affiliation{
Department of Applied Mathematics, \\
University of Western Ontario, \\
London, Ontario N6A 5B7, Canada}
\date{\today}

\preprint{KEK-TH-1345}
\preprint{UWO-TH-09/18}

\begin{abstract}
Recently, a new class of models describing the quark mass hierarchy has been
introduced. In this class, while the $t$ quark plays a minor role in
electroweak symmetry breaking (EWSB), it is crucial in providing the quark 
mass hierarchy. In this paper, we analyze the
dynamics of a particular model in this class, in which the 
$b'$ and $t'$ quarks of the 
fourth family are mostly responsible for dynamical EWSB. The low energy
effective theory in this model is derived. 
It has a clear
signature, a 2 + 1 structure of composite Higgs doublets: two nearly 
degenerate 
$\Phi_{b^{\prime}} \sim \bar{b^{\prime}}_{R}(t^{\prime},b^{\prime})_L$ and 
$\Phi_{t^{\prime}} \sim \bar{t^{\prime}}_{R}(t^{\prime},b^{\prime})_L$, 
and a heavier top-Higgs resonance $\Phi_t \sim \bar{t}_{R}(t,b)_L$.
The properties of these composites are described in detail, and 
it is shown  
that the model satisfies the electroweak precision data constraints. 
The signatures of these composites at the Large Hadron Collider 
are briefly discussed.
\end{abstract}

\maketitle

\section{Introduction}
\label{1}

The dynamics of electroweak symmetry breaking (EWSB) and fermion
(quark and lepton) mass hierarchy are the two central quests in the Large
Hadron Collider (LHC) program. In particular, it is noticeable 
that the LHC has a potential for discovering 
the fourth fermion family~\cite{AguilarSaavedra:2005pv}. 
The possibility of the existence of the latter
has been studied for a long time 
(for a review, see Ref.~\cite{Frampton:1999xi}).
It is noticeable that the
fourth family can play an important role
in B-CP asymmetries phenomena \cite{Hou:2006zza,Soni:2008bc}.

Since the mass bounds for the fourth family quarks $t'$ and $b'$ are 
of the order of the EWSB scale~\cite{pdg}, 
the Pagels-Stokar (PS) formula~\cite{PS} suggests that 
their contributions to the EWSB should not be small.
This leads to an idea of the dynamical EWSB scenario 
with the fourth family~\cite{4family,Simonov:2009hv}, which is 
an alternative version of the top 
quark condensate model 
\cite{MTY,Nambu,Bardeen:1989ds,footnote}.
Because the Yukawa couplings of the $t'$ and $b'$ quarks have 
the Landau pole around several TeV scale, it suggests that the
Higgs doublets 
$\Phi_{t^{\prime}} \sim 
\bar{t^{\prime}}_{R}(t^{\prime},b^{\prime})_L$ 
and 
$\Phi_{b^{\prime}} \sim \bar{b^{\prime}}_{R}(t^{\prime},b^{\prime})_L$
composed of them could be produced without fine tuning.

Although the top quark mass is obviously near the EWSB scale,
it apparently plays no leading role in the EWSB: the PS formula suggests 
that its contribution to the EWSB is around 10-20\%. On the other hand,
the $t$ quark might play an important role in the dynamics responsible 
for the quark 
mass hierarchy. Recently, 
utilizing dynamics considered in Ref.~\cite{Mendel:1991cx}
quite time ago, we introduced a new class of models
in which the top quark plays just such a role \cite{Hashimoto:2009xi}. 
The main two
features of these models are a) the presence of strong (although subcritical)
horizontal diagonal interactions for the $t$ quark, and b) horizontal 
flavor-changing neutral interactions between different families. 
Together
with the assumption that the dynamics primarily responsible for the EWSB 
leads to the mass spectrum of quarks with no (or weak) isospin violation, and
with the masses of the order of the observed masses of the down-type quarks,
these features allow to reproduce
the quark mass hierarchy and essential characteristics of 
the 
Cabibbo-Kobayashi-Maskawa (CKM) matrix~\cite{Hashimoto:2009xi}.

This approach 
can be implemented in the models with different EWSB scenarios.
Its signature is the existence of an additional top-Higgs resonance doublet
$\Phi_t$ composed of the quarks and antiquarks of the 3rd family,
$\Phi_t \sim \bar{t}_{R}(t,b)_L$. In the case of the 
dynamical EWSB scenario with the fourth family,
the top-Higgs $\Phi_t$ is heavier than the 
$\Phi_{t^{\prime}}$ and $\Phi_{b^{\prime}}$ composites 
\cite{Hashimoto:2009xi}. For simplicity, in Ref.~\cite{Hashimoto:2009xi} 
we considered only the case when
the $\Phi_t$ mass is ultraheavy and it decouples from TeV
dynamics. However, in general, this is not the case, 
and the $\Phi_t$ can be detectable at the LHC. 
This leads to a model with three 
Higgs doublets. Actually, because the 
$\Phi_{b} \sim \bar{b}_{R}(t,b)_L$ composite, and those ones connected
with the lighter $c, s, u$ and $d$ quarks, are necessarily 
ultraheavy and decouple in this scenario \cite{Hashimoto:2009xi},
and because there is an approximate $SU(2)_{R4}$ symmetry between
$t'_R$ and $b'_R$ quarks,
it would be appropriate to call it the 2 + 1 composite Higgs model.
In this paper, we will study such dynamics.

As for the fourth family leptons,
we assume that their masses are around 100~GeV~\cite{pdg},
and thus their contributions to the EWSB are smaller than 
that of the top quark.
For the dynamics with very heavy fourth family leptons,
and thereby with a lepton condensation, 
one needs to use, say, a five Higgs model. 
Also, the Majorana condensation of the right-handed neutrinos
should be reanalyzed in that case.
This possibility will be considered elsewhere.

The paper is organized as follows. In Sec. \ref{2}, we
describe the model. The qualitative features of its low energy 
effective theory are discussed in Sec. \ref{L-eff}. In 
Sec. \ref{4}, the results of the numerical analysis of the 
renormalization group equations are presented and the properties
of the composite Higgs bosons are described. 
The structure of the CKM matrix  
and flavor-changing-neutral interactions are
discussed in Sec. \ref{5}. 
In Sec. \ref{6}, we summarize the main results of the paper.
In Appendixes A-C, useful formulas used in the main text are
derived.

\section{Model}
\label{2}

We will utilize a Nambu-Jona-Lasinio (NJL) type model to describe
the dynamics with the $2 + 1$ Higgs doublets composed of 
the third and fourth family quarks. Its Lagrangian density has the 
following form:
\begin{equation}
    {\cal L} = {\cal L}_g + {\cal L}_f + {\cal L}_{\rm NJL},
\end{equation}
where ${\cal L}_g$ is the Lagrangian density for the Standard Model (SM)
gauge bosons,
the fermion kinetic term is
\begin{equation}
 {\cal L}_f \equiv 
   \sum_{i=3,4} \bar{\psi}^{(i)}_L i\fsl{D}\psi^{(i)}_L
 + \sum_{i=3,4} \bar{u}^{(i)}_R i\fsl{D}u^{(i)}_R
 + \sum_{i=3,4} \bar{d}^{(i)}_R i\fsl{D}d^{(i)}_R,
\end{equation}
and the NJL interactions are described by
\begin{eqnarray}
  {\cal L}_{\rm NJL} &=& 
   G_{t'}(\bar{\psi}_L^{(4)}t'_R)(\bar{t}'_R \psi_L^{(4)})
 + G_{b'}(\bar{\psi}_L^{(4)}b'_R)(\bar{b}'_R \psi_L^{(4)})
 + G_{t}(\bar{\psi}_L^{(3)}t_R)(\bar{t}_R \psi_L^{(3)}) \nonumber \\
&&
 + G_{t'b'}(\bar{\psi}_L^{(4)}t'_R)(\bar{b}'_R{}^c i\tau_2 (\psi_L^{(4)})^c )
 + G_{t't}(\bar{\psi}_L^{(4)}t'_R)(\bar{t}_R \psi_L^{(3)})
 + G_{b't}(\bar{\psi}_L^{(3)}t_R)(\bar{b}'_R{}^c i\tau_2 (\psi_L^{(4)})^c )
 + \mbox{(h.c.)}. 
\end{eqnarray}
Here $\psi_L^{(i)}$ denotes 
the weak doublet quarks from the $i$-th family,
and $u_R^{(i)}$ and $d_R^{(i)}$ represent 
the right-handed up- and down-type quarks.

It is useful to rewrite this theory in an equivalent
form by introducing auxiliary fields,
$\Phi_{t'}^{(0)}$, $\Phi_{b'}^{(0)}$, $\Phi_{t}^{(0)}$: 
\begin{equation}
    {\cal L} = {\cal L}_f + {\cal L}_g + {\cal L}_{\rm aux},
\end{equation}
where
\begin{eqnarray}
 - {\cal L}_{\rm aux} &=& \bar{\psi}_L^{(4)}t'_R \Phi_{t'}^{(0)}
 + \bar{\psi}_L^{(4)}b'_R \tilde{\Phi}_{b'}^{(0)}
 + \bar{\psi}_L^{(3)}t_R \Phi_{t}^{(0)} + \mbox{(h.c.)} \nonumber \\
&&
 + M_{\Phi_{t'}^{(0)}}^2 
   \left(\left(\Phi_{t'}^{(0)}\right)^\dagger \Phi_{t'}^{(0)}\right)
 + M_{\Phi_{b'}^{(0)}}^2 
   \left(\left(\Phi_{b'}^{(0)}\right)^\dagger \Phi_{b'}^{(0)}\right)
 + M_{\Phi_{t}^{(0)}}^2 
   \left(\left(\Phi_{t}^{(0)}\right)^\dagger \Phi_{t}^{(0)}\right)
   \nonumber \\
&&
 + M_{\Phi_{t'}^{(0)}\Phi_{b'}^{(0)}}^2 
   \left(\left(\Phi_{t'}^{(0)}\right)^\dagger \Phi_{b'}^{(0)}\right)
 + M_{\Phi_{t'}^{(0)}\Phi_{t}^{(0)}}^2 
   \left(\left(\Phi_{t'}^{(0)}\right)^\dagger \Phi_{t}^{(0)}\right)
 + M_{\Phi_{b'}^{(0)}\Phi_{t}^{(0)}}^2 
   \left(\left(\Phi_{b'}^{(0)}\right)^\dagger \Phi_{t}^{(0)}\right)
 + \mbox{(h.c.)} , 
\end{eqnarray}
with
\begin{equation}
 \left(\begin{array}{ccc}
    M_{\Phi_{t'}^{(0)}}^2 & M_{\Phi_{t'}^{(0)}\Phi_{b'}^{(0)}}^2 &
    M_{\Phi_{t'}^{(0)}\Phi_{t}^{(0)}}^2 \\
    M_{\Phi_{t'}^{(0)}\Phi_{b'}^{(0)}}^2 & M_{\Phi_{b'}^{(0)}}^2 &
    M_{\Phi_{b'}^{(0)}\Phi_{t}^{(0)}}^2 \\
    M_{\Phi_{t'}^{(0)}\Phi_{t}^{(0)}}^2 & M_{\Phi_{b'}^{(0)}\Phi_{t}^{(0)}}^2 &
    M_{\Phi_{t}^{(0)}}^2
  \end{array}\right)
  = \left(
    \begin{array}{ccc}
      G_{t'} & G_{t'b'} & G_{t't} \\
      G_{t'b'} & G_{b'} & G_{b't} \\
      G_{t't} & G_{b't} & G_t 
    \end{array}\right)^{-1} \, ,
\end{equation}
and $\tilde{\Phi}_{b'}^{(0)} \equiv -i\tau_2(\Phi_{b'}^{(0)})^*$.
The following remark is in order.
If we added the Yukawa mixing terms, 
they could be erased by redefining the composite Higgs fields. 
For example, for the mixing term $\bar{\psi}_L^{(3)}t_R \Phi_{t'}^{(0)}$,
the redefinition would be
$\varphi_{t}^{(0)} = \Phi_{t}^{(0)} + \Phi_{t'}^{(0)}$, 
$\varphi_{t'}^{(0)} = \Phi_{t'}^{(0)}$.
Such non-unitary (but invertible) transformations are allowed
because there are no canonical kinetic terms for the auxiliary 
fields in ${\cal L}$.

As was shown in Ref. \cite{Hashimoto:2009xi}, 
the diagonal parts of the NJL interactions,
$G_{t'}$, $G_{b'}$ and $G_{t}$, can be generated from 
the topcolor interactions~\cite{Hill:2002ap}.
In this case, the scales for the dimensionful NJL parameters 
$G_{t'} \simeq G_{b'}$ and $G_t$ are connected with 
the coloron masses, $\Lambda^{(4)}$ and $\Lambda^{(3)}$, respectively.
The mixing term $G_{t't}$
can be generated by a flavor-changing-neutral (FCN) interaction,
$t'$-$t$-$\Lambda^{(34)}$~\cite{Hashimoto:2009xi}.
On the other hand, $G_{t'b'}$
may be connected with topcolor instantons~\cite{Hill:2002ap}.
In the 2 + 1 composite Higgs model,
while the coupling
constants $G_{t'}$ and $G_{b'}$ are supercritical and responsible 
for EWSB, the $t$ quark coupling $G_{t}$ is subcritical, although
also strong \cite{Hashimoto:2009xi}. 

As to the $G_{b't}$ term, the situation is the following.
As far as $M_{\Phi_{t'}^{(0)}\Phi_{b'}^{(0)}}^2 \ne 0$ and 
$M_{\Phi_{t'}^{(0)}\Phi_{t}^{(0)}}^2 \ne 0$,
there do not appear Nambu-Goldstone (NG) bosons 
even if the $M_{\Phi_{b'}^{(0)}\Phi_{t}^{(0)}}^2$ term,
which is connected with $G_{b't}$, is ignored.
For example,
assuming $M_{\Phi_{b'}^{(0)}\Phi_{t}^{(0)}}^2=0$,
the Peccei-Quinn like $U(1)_A$ symmetry,
\begin{subequations}
\begin{eqnarray}
&& \psi_L^{(3)} \to e^{-i\theta_A} \psi_L^{(3)}, \quad 
   \psi_L^{(4)} \to e^{-i\theta_A} \psi_L^{(4)}, \quad 
  t'_R \to e^{i\theta_A} t'_R, \quad b'_R \to e^{i\theta_A} b'_R, \quad 
  t_R \to e^{i\theta_A} t_R, \\
&& \Phi_{t}^{(0)}  \to e^{-2i\theta_A} \Phi_{t}^{(0)}, \quad
   \Phi_{t'}^{(0)} \to e^{-2i\theta_A} \Phi_{t'}^{(0)}, \quad
   \Phi_{b'}^{(0)} \to e^{2i\theta_A} \Phi_{b'}^{(0)}, 
\end{eqnarray}
\end{subequations}
is explicitly broken by the Higgs mass mixing term
$M_{\Phi_{t'}^{(0)}\Phi_{b'}^{(0)}}^2 \ne 0$.
(Although the mixing term $M_{\Phi_{t'}^{(0)}\Phi_{t}^{(0)}}^2$
does not break this $U(1)_A$ symmetry, it is important: if both
$M_{\Phi_{t'}^{(0)}\Phi_{t}^{(0)}}^2$ and 
$M_{\Phi_{b'}^{(0)}\Phi_{t}^{(0)}}^2$equal
zero, a new global $U(1)$ symmetry appears.)
Therefore, it is safe to take
$M_{\Phi_{b'}^{(0)}\Phi_{t}^{(0)}}^2=0$.
Because of that, although 
we will keep the $G_{b't}$ and $M_{\Phi_{b'}^{(0)}\Phi_{t}^{(0)}}^2$ 
terms in a general
discussion for a while, they will be ignored 
in the numerical analysis.

\section{Dynamics in the low energy effective model: qualitative
features}
\label{L-eff}

The model introduced in the previous section
provides an approximate 
$2+1$ structure in the
Higgs quartic coupling sector in the low energy 
effective action.
Indeed, in the bubble approximation, while
the top-Higgs $\Phi_t$ couples only to $\psi_L^{(3)}$ and 
$t_R$, 
the composite $\Phi_{t'(b')}$  couples  
only to 
$\psi_L^{(4)}$ and $t'_R (b'_R)$, that leads to  
such a 2 + 1 structure. 
When we turn on the electroweak gauge interactions,
this structure breaks down.
The breaking effects are however suppressed, 
because the Yukawa couplings are much larger than 
the electroweak gauge ones.

In this section, we analyze the main characteristics of
the $2+1$ low energy effective
Higgs model, in particular,
the structure of its vacuum expectation values (VEV).
We also discuss the relations between 
the parameters of the initial NJL model (such as the NJL couplings, etc.)
and the observable ones.

In order to illustrate main qualitative features
of the effective model,
we will employ the bubble approximation in calculating 
its parameters (such as Yukawa and quartic couplings, etc.).
However, the structure of the action will be taken to be more general,
based on a numerical analysis
of the renormalization group equations (RGE's)
with the compositeness conditions \cite{Bardeen:1989ds},
which is performed in the next section.

\subsection{Low energy effective model}
\label{3a}

Since at low energy the composite Higgs fields develop 
kinetic terms, the Lagrangian density of the
low energy effective model is
\begin{equation}
  {\cal L} = {\cal L}_f + {\cal L}_g + {\cal L}_s + {\cal L}_y, 
\end{equation}
with
\begin{eqnarray}
 {\cal L}_s &=& |D_\mu \Phi_{b'}|^2 + |D_\mu \Phi_{t'}|^2
 + |D_\mu \Phi_{t}|^2 - V,
\end{eqnarray}
and
\begin{equation}
 - {\cal L}_y = y_{b'} \bar{\psi}_L^{(4)}b'_R \tilde{\Phi}_{b'}
 + y_{t'} \bar{\psi}_L^{(4)}t'_R \Phi_{t'}
 + y_t \bar{\psi}_L^{(3)}t_R \Phi_{t} + \mbox{(h.c.)} , 
\label{yukawa}
\end{equation}
where $V$ is the Higgs potential and
$\Phi_{t',b',t}$ are the renormalized Higgs fields.
Taking into account the renormalization group (RG) improved analysis, 
which will be presented in the next section, 
we study the following Higgs potential:
\begin{equation}
  V = V_2 + V_4,
\end{equation}
with
\begin{eqnarray}
  V_2 &=& M_{\Phi_{b'}}^2 (\Phi_{b'}^\dagger \Phi_{b'})
 + M_{\Phi_{t'}}^2 (\Phi_{t'}^\dagger \Phi_{t'})
 + M_{\Phi_{t}}^2 (\Phi_{t}^\dagger \Phi_{t})
   \nonumber \\
&&
 + M_{\Phi_{t'}\Phi_{b'}}^2 (\Phi_{t'}^\dagger \Phi_{b'})
 + M_{\Phi_{b'}\Phi_{t}}^2 (\Phi_{b'}^\dagger \Phi_{t})
 + M_{\Phi_{t'}\Phi_{t}}^2 (\Phi_{t'}^\dagger \Phi_{t})
 + \mbox{(h.c.)}, \\
  V_4 &=& \lambda_1 (\Phi_{b'}^\dagger \Phi_{b'})^2
  + \lambda_2 (\Phi_{t'}^\dagger \Phi_{t'})^2
  + \lambda_3 (\Phi_{b'}^\dagger \Phi_{b'})(\Phi_{t'}^\dagger \Phi_{t'})
  + \lambda_4 |\Phi_{b'}^\dagger \Phi_{t'}|^2
  + \frac{1}{2}
    \bigg[\,\lambda_5 (\Phi_{b'}^\dagger \Phi_{t'})^2 + \mbox{(h.c.)}\,\bigg]
  + \lambda_t (\Phi_{t}^\dagger \Phi_{t})^2 \, .
\end{eqnarray}
While $M_{\Phi_{b'}}^2$ and $M_{\Phi_{t'}}^2$ are negative, the mass
square $M_{\Phi_{t}}^2$ is positive, which reflects a subcritical
dynamics of the $t$ quark. The top-Higgs $\Phi_{t}$ acquires a vacuum
expectation value only due to its mixing with $\Phi_{t'}$ (as was already
indicated above, we assume that its mixing with $\Phi_{b'}$ is negligible).

The bubble approximation yields
the following Yukawa couplings 
\begin{eqnarray}
  y_{q'}(\mu) & \equiv & y_{t'}(\mu) = y_{b'}(\mu) =
  \left(\frac{N}{16\pi^2}\ln \frac{(\Lambda^{(4)})^2}{\mu^2}\right)^{-1/2},\\
  y_{t}(\mu) &=&
  \left(\frac{N}{16\pi^2}\ln \frac{(\Lambda^{(3)})^2}{\mu^2}\right)^{-1/2},
\end{eqnarray}
the Higgs mass terms,
\begin{eqnarray}
  M_{\Phi_{t'}}^2(\mu) &=& 
  y_{t'}^2 \left[\,M_{\Phi_{t'}^{(0)}}^2
                - \frac{N}{8\pi^2}((\Lambda^{(4)})^2 - \mu^2)\,\right],\\
  M_{\Phi_{b'}}^2(\mu) &=& 
  y_{b'}^2 \left[\,M_{\Phi_{b'}^{(0)}}^2
                - \frac{N}{8\pi^2}((\Lambda^{(4)})^2 - \mu^2)\,\right],\\
  M_{\Phi_{t}}^2(\mu) &=& 
  y_{t}^2 \left[\,M_{\Phi_{t}^{(0)}}^2
                - \frac{N}{8\pi^2}((\Lambda^{(3)})^2 - \mu^2)\,\right], \\
  M_{\Phi_{t'}\Phi_{b'}}^2 &=& y_{t'} y_{b'} M_{\Phi_{t'}^{(0)}\Phi_{b'}^{(0)}}^2, \\
  M_{\Phi_{t'}\Phi_{t}}^2 &=& y_{t'} y_{t} M_{\Phi_{t'}^{(0)}\Phi_{t}^{(0)}}^2, \\
  M_{\Phi_{b'}\Phi_{t}}^2 &=& y_{t} y_{b'} M_{\Phi_{b'}^{(0)}\Phi_{t}^{(0)}}^2, 
\end{eqnarray}
and the Higgs quartic couplings,
\begin{equation}
  \lambda_1 = \lambda_2 = \frac{\lambda_3}{2}=-\frac{\lambda_4}{2}
 = y_{q'}^2,  \quad \lambda_5 = 0 , 
 \label{lam}
\end{equation}
\begin{equation}
  \lambda_t = y_{t}^2 ,  
\end{equation}
where $N(=3)$ denotes the color number,
$\mu$ is a renormalization scale, and 
$\Lambda^{(3),(4)}$ are the composite scales for the top and 
the fourth family quarks, respectively.
For details, see Appendix~\ref{bubble-app}.

While the structure of the mass term part $V_2$ is general for 
three Higgs doublet models, 
the $V_4$ part is presented as the sum of the potential for 
the two Higgs doublets $\Phi_{t'}$ and $\Phi_{b'}$
and that for the doublet $\Phi_{t}$, i.e., it reflects 
the 2 + 1 structure of the present model.
For the most general three Higgs potential, see Appendix~\ref{3higgs}.

As far as we ignore the electroweak (EW) gauge interactions,
the terms breaking the $(2+1)$-Higgs structure,
such as $(\Phi_{t'}^\dagger \Phi_{t'})(\Phi_{t}^\dagger \Phi_{t})$,
are not generated by the one-loop diagrams.
The $(2+1)$-Higgs approximation should work well even 
in the numerical analysis:
We expect that the errors connected with this approximation
is at most around few \%,
and hence they are less than a 10\% level
uncertainty of nonperturbative effects,
which will be discussed in the next section.
Note that while the $1/N$-leading approximation,
including the QCD effects, is qualitatively reasonable, 
it is not good quantitatively, with errors
around 30\% level.

In passing, because the NJL model is used, 
eight-Fermi interactions, such as 
$|\bar{\psi}_L^{(4)}t'_R|^2 |\bar{\psi}_L^{(3)} t_R|^2$,
are ignored in the present approach. 
This point is also important for keeping the $(2+1)$-Higgs structure.

\subsection{The structure of the vacuum expectation values}
\label{3b}

Let us analyze the VEV structure
and the mass spectrum of 
the fourth family quarks and the Higgs bosons.

We define the components of the Higgs fields by
\begin{equation}
 \Phi_X = 
  \left(
   \begin{array}{c}
    \frac{1}{\sqrt{2}} (v_X + h_X-i z_X) \\ -\omega_X^-
    \end{array}
  \right) ,   \quad \tilde{\Phi}_X \equiv -i\tau_2\Phi_X^*,
\label{comp}
\end{equation}
where $X=b',t',t$.
Note that the relation
\begin{equation}
  v^2 = v_{b'}^2 + v_{t'}^2 + v_t^2,
  \label{vev}
\end{equation}
holds, where $v\simeq$246 GeV.
It is convenient to introduce the ratio of VEVs,
\begin{equation}
  \tan\beta_4 \equiv \frac{v_{t'}}{v_{b'}}, \quad
  \tan\beta_{34}\equiv \frac{v_t}{\sqrt{v_{t'}^2+v_{b'}^2}},
\end{equation}
i.e.,
\begin{eqnarray}
  v_{b'} &=& v \cos\beta_4\cos\beta_{34}, \\
  v_{t'} &=& v \sin\beta_4\cos\beta_{34}, \\
  v_{t} &=& v \sin\beta_{34} \, . 
\end{eqnarray}
The notations $s_{\beta_4} \equiv \sin\beta_4$, 
$s_{\beta_{34}} \equiv \sin\beta_{34}$, etc., will be used.
The quark masses are (compare with Eq. (\ref{yukawa})):
\begin{eqnarray}
  m_{b'} &=& \frac{v_{b'}}{\sqrt{2}}\;y_{b'}(\mu=m_{b'}), \\
  m_{t'} &=& \frac{v_{t'}}{\sqrt{2}}\;y_{t'}(\mu=m_{t'}), \\
  m_{t}  &=& \frac{v_{t}}{\sqrt{2}} \;y_{t}(\mu=m_{t})\,. 
\end{eqnarray}
Since we expect $\Lambda^{(4)} \sim \Lambda^{(3)}$,
the Yukawa couplings are almost the same,
$y_{t'}(\mu=m_{t'}) \simeq y_{b'}(\mu=m_{b'}) \sim y_{t}(\mu=m_{t})$.
The $T$-parameter constraint suggests that
$m_{t'} \simeq m_{b'}$ is favorable, so that
the phenomenological condition $m_{t'} \simeq m_{b'} \gtrsim m_t$ 
requires $v_{t'} \simeq v_{b'} \gtrsim v_t$, i.e.,
\begin{equation}
  \tan\beta_4 \simeq 1, \quad \tan\beta_{34} \lesssim 1 \, .
\end{equation}
To obtain $\tan\beta_{34} \lesssim 1$, the subcritical 
dynamics for the $t$ quark, leading to
$M_{\Phi_{t}}^2 > 0$, is crucial~\cite{Hashimoto:2009xi}.

Let us analyze the VEV structure and how we can obtain 
the desirable solution.
The effective potential expressed through the VEVs
is given by
\begin{eqnarray}
  V_{\rm eff} &=& \frac{1}{2} M_{\Phi_{b'}}^2 v_{b'}^2
 + \frac{1}{2}M_{\Phi_{t'}}^2 v_{t'}^2
 + \frac{1}{2} M_{\Phi_{t}}^2 v_{t}^2
 + M_{\Phi_{t'}\Phi_{b'}}^2 v_{t'} v_{b'}
 + M_{\Phi_{b'}\Phi_{t}}^2 v_{b'} v_{t}
 + M_{\Phi_{t'}\Phi_{t}}^2 v_{t'} v_{t} \nonumber \\
&&
 + \frac{1}{4}\lambda_1 v_{b'}^4 + \frac{1}{4}\lambda_2 v_{t'}^4
 + \frac{1}{4}(\lambda_3 + \lambda_4 + \lambda_5) v_{t'}^2 v_{b'}^2
 + \frac{1}{4}\lambda_t v_t^4 ,
\end{eqnarray}
so that the stationary conditions are
\begin{eqnarray}
  \frac{\partial V_{\rm eff}}{\partial v_{b'}} &=&
    M_{\Phi_{b'}}^2 v_{b'} + M_{\Phi_{t'} \Phi_{b'}}^2 v_{t'}
  + M_{\Phi_{b'} \Phi_t}^2 v_t
  + \lambda_1 v_{b'}^3
  + \frac{1}{2}(\lambda_3 + \lambda_4 + \lambda_5) v_{b'} v_{t'}^2 =0, \\ 
  \frac{\partial V_{\rm eff}}{\partial v_{t'}} &=&
    M_{\Phi_{t'}}^2 v_{t'} + M_{\Phi_{t'} \Phi_{b'}}^2 v_{b'}
  + M_{\Phi_{t'} \Phi_t}^2 v_t
  + \lambda_2 v_{t'}^3
  + \frac{1}{2}(\lambda_3 + \lambda_4 + \lambda_5) v_{t'} v_{b'}^2 =0, \\
  \frac{\partial V_{\rm eff}}{\partial v_t} &=&
   M_{\Phi_t}^2 v_t + M_{\Phi_{t'} \Phi_t}^2 v_{t'}
  + M_{\Phi_{b'} \Phi_t}^2 v_{b'}
  + \lambda_t v_t^3 =0 \, .
\end{eqnarray}
In order to obtain the approximate solution with
$v_{t'} \simeq v_{b'} \gtrsim v_t$, we assume 
\begin{equation}
  |M_{\Phi_{t'}}^2| \approx |M_{\Phi_{b'}}^2| \gtrsim 
  \frac{v_t}{v_{t'}}|M_{\Phi_{t'} \Phi_t}^2|, 
  \frac{v_t}{v_{b'}}|M_{\Phi_{b'} \Phi_t}^2| \, . 
\end{equation}
These assumptions are easily satisfied in our dynamical model.
If we further impose 
\begin{equation}
  M_{\Phi_t}^2 \gg \lambda_t v_t^2, 
\end{equation}
and
\begin{equation}
  |M_{\Phi_{t'} \Phi_t}^2| v_{t'} \gg \lambda_t v_t^3,
\end{equation}
the solution is approximately given by
\begin{eqnarray}
&& \left[\lambda_1 + \frac{1}{2}(\lambda_3+\lambda_4 + \lambda_5)
         \tan^2\beta_4\right] v_{b'}^2
 \simeq - M_{\Phi_{b'}}^2 - M_{\Phi_{t'} \Phi_{b'}}^2 \tan\beta_4, \\
&& \left[\lambda_2 + \frac{1}{2}(\lambda_3+\lambda_4 + \lambda_5)
         \cot^2\beta_4\right] v_{t'}^2
 \simeq - M_{\Phi_{t'}}^2 - M_{\Phi_{t'} \Phi_{b'}}^2 \cot\beta_4, \\
&& v_t \simeq \frac{-M_{\Phi_{t'} \Phi_t}^2}{M_{\Phi_t}^2}v_{t'}
           +\frac{-M_{\Phi_{b'} \Phi_t}^2}{M_{\Phi_t}^2}v_{b'}\,.
 \label{app-st}
\end{eqnarray}
The last equation essentially determines $\tan\beta_{34}$.

\subsection{Mass spectrum of the Higgs bosons}
\label{3c}

We now analyze the mass spectrum of the Higgs bosons.

The formulas for the masses of the CP even Higgs bosons are 
quite complicated, because they are the eigenvalues of
the $3 \times 3$ matrices.
Even for the CP odd and charged Higgs bosons, 
the mass formulas are still not quite simple
(for the analytic formulas, see Appendix~\ref{analy-mass}).
In order to make the physical meaning of the dynamics
more transparent, here
we will consider approximate and useful expressions.

The $T$-parameter constraint suggests
$\tan\beta_4 \simeq 1$. At the zeroth approximation,
we may take exactly $\tan\beta_4=1$.
As was pointed out in Sec. \ref{2}, we may further assume
$M_{\Phi_{b'} \Phi_t}^2 \approx 0$.
We also find $\lambda_5 = 0$ (see Sec. \ref{3a} above).

The mass of the charged top-Higgs boson, 
which mainly couples to the top and bottom, should be
constrained by $R_b$ and, therefore, should be rather heavy.
We thus conclude that each of the heaviest CP even, CP odd and 
charged Higgs bosons are mainly provided by the 
top-Higgs doublet $\Phi_t$.

Then the mass eigenvalues are approximately given by
\begin{eqnarray}
  M_{A_1}^2 & \simeq &
  -2M_{\Phi_{t'} \Phi_{b'}}^2 (1-\tan^2\beta_{34}), 
  \label{app-ma1} \\
  M_{A_2}^2 & \simeq & M_{\Phi_{t}}^2(1+2\tan^2\beta_{34})
  + M_{A_1}^2\tan^2\beta_{34},
  \label{app-ma2} \\
  M_{H_1^\pm}^2 & \simeq & M_{A_1}^2
  -\frac{1}{2}\lambda_4 v^2 c_{\beta_{34}}^2 (1-\tan^2\beta_{34}), 
  \label{app-mhpm1} \\
  M_{H_2^\pm}^2 & \simeq & M_{A_2}^2
  -\frac{1}{2} \lambda_4 v^2 s^2_{\beta_{34}} ,
  \label{app-mhpm2}
\end{eqnarray}
up to ${\cal O}(\tan^2\beta_{34})$.
Here for the CP odd Higgs bosons and for the charged Higgs bosons, 
we defined
$M_{A_1} \leq M_{A_2}$ and  
$M_{H_1^\pm} \leq M_{H_2^\pm}$, respectively.
For the CP even Higgs bosons, we 
defined $M_{H_1} \leq M_{H_2} \leq M_{H_3}$.
As was indicated above,
the heavy Higgs bosons, $H_2^\pm, A_2$, and $H_3$, 
consist mainly of the components of the top-Higgs $\Phi_t$.

The stationary condition (\ref{app-st}) approximately read
\begin{equation}
  \frac{-M_{\Phi_{t'} \Phi_t}^2}{M_{\Phi_t}^2} \approx
  \sqrt{2}\tan\beta_{34},
\end{equation}
where we took $M_{\Phi_{b'} \Phi_t}^2 = 0$.
By using Eq.~(\ref{lam}), Eqs.~(\ref{app-ma1})--(\ref{app-mhpm2}),
$v^2 c_{34}^2=v_{t'}^2+v_{b'}^2$, and 
$-\lambda_4 v_{t'(b')}^2 = 4 m_{t'(b')}^2$ in the bubble approximation,
we also find the charged Higgs masses as
\begin{eqnarray}
  M_{H_1^\pm}^2 & \approx &
  M_{A_1}^2 + 2(m_{t'}^2+m_{b'}^2)(1-\tan^2\beta_{34}), 
  \label{app2-mhpm1} \\
  M_{H_2^\pm}^2 & \approx &
  M_{A_2}^2 + 2(m_{t'}^2+m_{b'}^2)\tan^2\beta_{34} \, . 
  \label{app2-mhpm2}
\end{eqnarray}
The upper bound of $M_{A_1}$ for a given value of $M_{A_2}$
is discussed in Appendix~\ref{analy-mass}.

There are eight parameters in the initial NJL model: 
six NJL couplings and two composite scales, $\Lambda^{(3,4)}$.
As we discussed above, these parameters are closely connected with
physical observables.
The values of $\Lambda^{(3,4)}$ determine the Yukawa couplings.
Then, by using the experimental value of $m_t$, we can find $v_t$.
Fixing the value of $\tan\beta_4$,
we can determine $v_{t'}$ and $v_{b'}$  through Eq.~(\ref{vev}),
and thereby can express $m_{t'}$ and $m_{b'}$ through 
the Yukawa couplings.
The masses $M_{\Phi_{t'} \Phi_{b'}}^2$ and 
$M_{\Phi_{t}}^2$ are connected with 
$M_{A_1}^2$ and $M_{A_2}^2$, respectively.
The value of $M_{\Phi_{t'} \Phi_{t}}^2/M_{\Phi_{t}}^2$ is
approximately given by $v_t/v_{t'}$, if we assume
$M_{\Phi_{b'} \Phi_{t}}^2 \approx 0$, 
as we already did.

In summary, it is convenient to take the following eight parameters 
instead of the original theoretical ones:
\begin{equation}
  v(=\mbox{246 GeV}), \quad m_t(=\mbox{171.2 GeV}), \quad
  \tan\beta_4(\simeq 1), \quad 
  M_{A_1}, \quad M_{A_2}, \quad 
  \Lambda^{(3)}, \quad \Lambda^{(4)}, \quad 
  M_{\Phi_{b'} \Phi_{t}}^2 (\approx 0)\,.
\end{equation}
In the next section, we will perform a numerical analysis.

\section{Numerical analysis}
\label{4}

The analysis in the previous section was somewhat schematic.
In this section, in order to describe the dynamics in the model 
more precisely, we will employ the RGE's with the compositeness 
conditions~ 
\cite{Bardeen:1989ds,Luty:1990bg}:
\begin{equation}
  y_{t'}^2(\mu=\Lambda^{(4)})=\infty, \quad
  y_{b'}^2(\mu=\Lambda^{(4)})=\infty, \quad
  y_{t}^2(\mu=\Lambda^{(3)})=\infty, 
\end{equation}
and
\begin{equation}
  \frac{\lambda_1}{y_{b'}^4}\bigg|_{\mu=\Lambda^{(4)}}=0, \quad
  \frac{\lambda_2}{y_{t'}^4}\bigg|_{\mu=\Lambda^{(4)}}=0, \quad
  \frac{\lambda_3}{y_{b'}^2 y_{t'}^2}\bigg|_{\mu=\Lambda^{(4)}}=0, \quad
  \frac{\lambda_4}{y_{b'}^2 y_{t'}^2}\bigg|_{\mu=\Lambda^{(4)}}=0, \quad
  \frac{\lambda_t}{y_{t}^4}\bigg|_{\mu=\Lambda^{(3)}}=0 \, . 
\end{equation}

The RGE's are similar to those for the 
two Higgs doublet model (THDM) type II~\cite{Hill:1985tg}.
For consistency with the $(2+1)$-Higgs structure, 
we ignore the one-loop effects of the EW interactions,
which should be tiny.
On the other hand, although the Higgs loop effects are
of the $1/N$-subleading order, we incorporate them, 
because they are numerically relevant.

The RGE for the QCD coupling is given by
\begin{equation}
(16 \pi^2) \mu \frac{\partial }{\partial \mu} g_3 = - c_3 g_3^3,  
\qquad c_3 = 11 - \frac{4}{3}N_g,
\end{equation} 
where $N_g$ denotes the number of generations (families).
The RGE's for Yukawa couplings are 
\begin{subequations}
\begin{eqnarray}
(16 \pi^2) \mu \frac{\partial }{\partial \mu} y_{b'} &=&
 - 8g_3^2 y_{b'} + \frac{9}{2}y_{b'}^3
 + \frac{1}{2}y_{t'}^2 y_{b'}, \\
(16 \pi^2) \mu \frac{\partial }{\partial \mu} y_{t'} &=&
 - 8g_3^2 y_{t'} + \frac{9}{2}y_{t'}^3
 + \frac{1}{2}y_{b'}^2 y_{t'}, \\
(16 \pi^2) \mu \frac{\partial }{\partial \mu} y_{t} &=&
 - 8g_3^2 y_t + \frac{9}{2}y_t^3,
\end{eqnarray}
\end{subequations}
where we ignored the bottom Yukawa coupling $y_b$ and
the EW loop effects in order to keep the $(2+1)$-Higgs structure.
On the other hand, the RGE's for the Higgs quartic self-couplings are 
\begin{eqnarray}
(16 \pi^2) \mu \frac{\partial }{\partial \mu} \lambda_1 = &&
   24\lambda_1^2+2\lambda_3^2+2\lambda_3\lambda_4+\lambda_4^2
  +12\lambda_1y_{b'}^2-6y_{b'}^4, \\
(16 \pi^2) \mu \frac{\partial }{\partial \mu} \lambda_2 = &&
    24\lambda_2^2+2\lambda_3^2+2\lambda_3\lambda_4+\lambda_4^2
   +12\lambda_2 y_{t'}^2 - 6y_{t'}^4, \\
(16 \pi^2) \mu \frac{\partial }{\partial \mu} \lambda_3 = &&
    2(\lambda_1+\lambda_2)(6\lambda_3+2\lambda_4)+4\lambda_3^2
  +2\lambda_4^2 +6\lambda_3(y_{b'}^2+y_{t'}^2)-12y_{b'}^2y_{t'}^2, \\
(16 \pi^2) \mu \frac{\partial }{\partial \mu} \lambda_4 = &&
   4(\lambda_1+\lambda_2)\lambda_4
  +4(2\lambda_3+\lambda_4)\lambda_4
  +6\lambda_4(y_{b'}^2+y_{t'}^2) + 12y_{b'}^2y_{t'}^2, \\
(16 \pi^2) \mu \frac{\partial }{\partial \mu} \lambda_5 = &&
    \lambda_5\bigg[\,4(\lambda_1+\lambda_2)+8\lambda_3+12\lambda_4
   +6(y_{t'}^2 + y_{b'}^2)\,\bigg], \\
(16 \pi^2) \mu \frac{\partial }{\partial \mu} \lambda_t = &&
    24\lambda_t^2 + 12\lambda_t y_t^2-6y_t^4,  
\end{eqnarray}
where we ignored the EW loop effects.
Note that the coupling
constants $\lambda_1$ and $\lambda_2$ that we use are 
twice larger than those in Ref.~\cite{Hill:1985tg}.
In our model, we find $\lambda_5=0$.

Since we impose the same compositeness condition for 
$t'$ and $b'$, 
and because the RGE's for $y_{t'}$ and $y_{b'}$ are the same, 
the $SU(2)_{R4}$ symmetry, which is the symmetry between 
$t'_R$ and $b'_R$, is exact for both the Yukawa and Higgs quartic 
couplings, as far as the EW interactions are ignored.
The $SU(2)_{R4}$ breaking effects appear only from the 
Higgs mass mixing terms.
This leads to  $v_{t'} \ne v_{b'}$ in general, and thereby
the mass difference between the $t'$ and $b'$ quarks
can arise.

For the numerical calculations, 
we vary $M_{A_1}$, $M_{A_2}$, $\Lambda^{(3)}$, $\Lambda^{(4)}$, 
$\tan\beta_4$, 
and, as an input, use $v = 246$ GeV and
the $\overline{\rm MS}$-mass $m_t=161.8$ GeV. The latter 
corresponds to the pole mass $M_t=171.2$ GeV~\cite{pdg}.
We also use the QCD coupling constant $\alpha_3(M_Z)=0.1176$~\cite{pdg}.
As for $M_{\Phi_{b'} \Phi_{t}}^2$,
we fix $M_{\Phi_{b'} \Phi_{t}}^2 = 0$.
Numerically, it is consistent with $G_{b' t} \approx 0$.

\begin{figure}[t]
  \begin{flushleft}
  \includegraphics{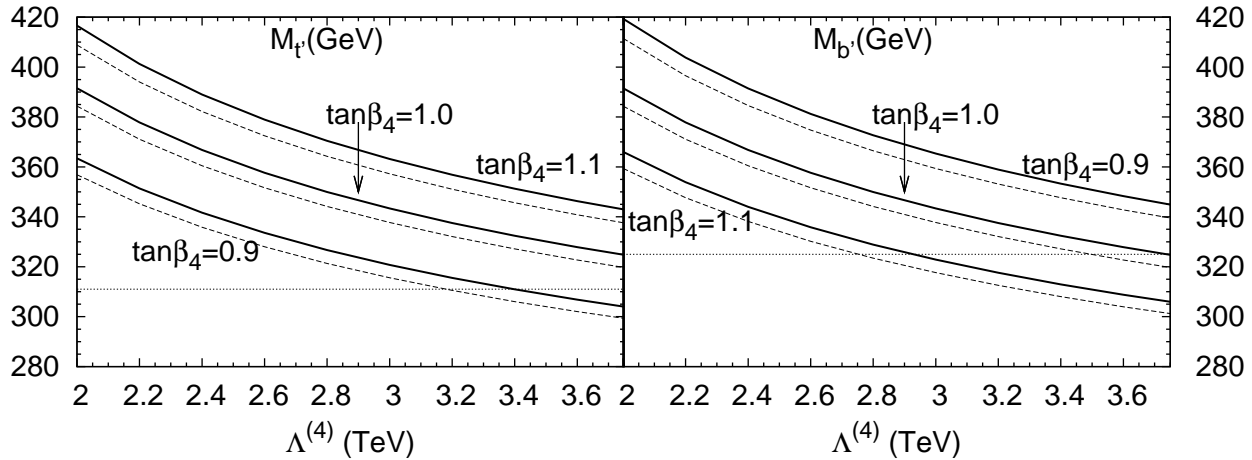}
  \end{flushleft}
  \caption{$M_{t'}$ and $M_{b'}$. 
  The bold and dashed curves are for
  $\Lambda^{(3)}/\Lambda^{(4)}=1,2$, respectively.
  The dotted lines correspond to the lower bounds for 
  the masses of $t'$ and $b'$ at 95\% C.L., 
  $M_{t'} > 311$~GeV and $M_{b'} > 325$~GeV~\cite{CDF-bound}.
  \label{mtp-mbp}}
\end{figure}

\begin{figure}[t]
  \begin{flushleft}
  \includegraphics{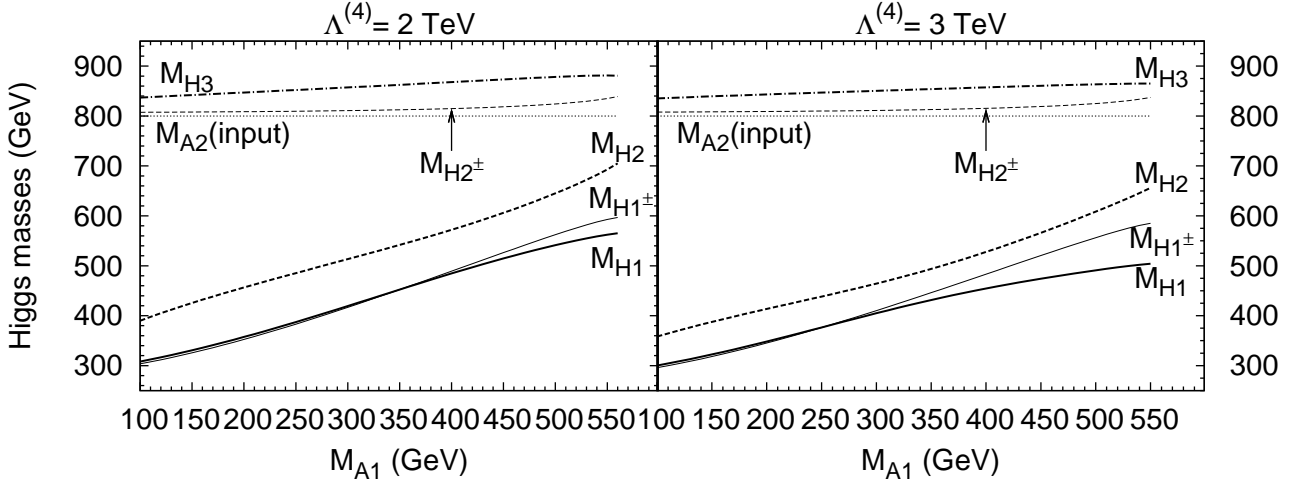}
  \end{flushleft}
  \caption{Mass spectrum of the Higgs bosons for
  $\Lambda^{(4)}=2,3$ TeV .
  We took $\Lambda^{(3)}/\Lambda^{(4)}=1.5$ and $\tan\beta_4=1$.
  $M_{A_2}=800$ GeV is the input. 
  The 2 + 1 structure in the model is clearly 
   manifested.
  \label{ma1-ma2}}
\end{figure}

\begin{figure}[t]
  \begin{flushleft}
  \includegraphics{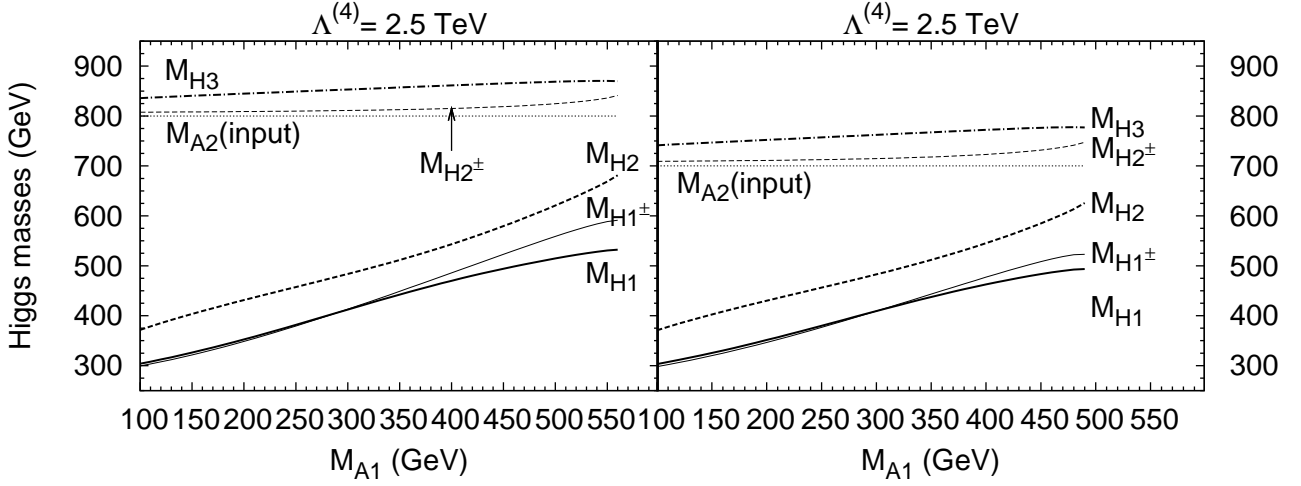}
  \end{flushleft}
  \caption{Mass spectrum of the Higgs bosons for
  $\Lambda^{(4)}=2.5$ TeV .
  We took $\Lambda^{(3)}/\Lambda^{(4)}=1.5$ and $\tan\beta_4=1$.
  The CP odd Higgs masses $M_{A_2}=800,700$ GeV are also the inputs.
  The 2 + 1 structure in the model is clearly manifested.
  \label{ma2-dep}}
\end{figure}

The results are illustrated in Figs.~\ref{mtp-mbp}--\ref{ma2-dep}.
The masses of $t'$ and $b'$ are essentially determined by
the value of $\Lambda^{(4)}$, 
where we converted the $\overline{\rm MS}$-masses $m_{t'}$ and 
$m_{b'}$ to the on-shell ones, 
$M_{t'(b')}=m_{t'(b')}[1+4\alpha_s/(3\pi)]$. 
As is seen in Fig. \ref{mtp-mbp},
their dependence on $\Lambda^{(3)}/\Lambda^{(4)} (=1-2)$
is mild.
When we vary $\tan\beta_4$ in the interval $0.9\mbox{--}1.1$,
the variations of $M_{t'}$ and $M_{b'}$ are up to 10\% 
(see Fig.~\ref{mtp-mbp}).

The Higgs masses are sensitive to the value of $\Lambda^{(4)}$
(see Fig.~\ref{ma1-ma2}), 
while their sensitivity to $\Lambda^{(3)}/\Lambda^{(4)} (=1-2)$
is low.
Note also that the Higgs mass dependence on $\tan\beta_4$
is mild, at most 5\% for $\tan\beta_4=0.9\mbox{--}1.1$,
$\Lambda^{(4)}=2\mbox{--}10$ TeV, and $\Lambda^{(3)}/\Lambda^{(4)}=1.5$.

It is noticeable
that the masses of the $H_2^\pm$ and $H_3$ Higgs bosons
are close and correlate with the mass of the $A_2$ boson, as shown in 
Fig.~\ref{ma2-dep}. This point agrees with that we identified these 
heaviest bosons mostly with the top-Higgs doublet $\Phi_t$:
it reflects a subcritical dynamics of the $t$ quark. 
Last but not least, Figs.~\ref{ma1-ma2} and 
\ref{ma2-dep} clearly illustrate the 2 + 1 structure of the model.

Since at the compositeness scale the Yukawa couplings go to infinity,
there could in principle be uncontrollable nonperturbative effects.
In order to estimate them, we studied the RGE's with
relaxed compositeness conditions:
\begin{equation}
  y_{t'}^2(\mu=\Lambda^{(4)})=y_\Lambda^2, \quad
  y_{b'}^2(\mu=\Lambda^{(4)})=y_\Lambda^2, \quad
  y_{t}^2(\mu=\Lambda^{(3)})=y_\Lambda^2, \quad y_\Lambda^2 < \infty\,.
\end{equation}
For concreteness, we took $y_\Lambda^2=25$.
It was found that such nonperturbative effects are 
around ${\cal O}(\mbox{10 \%})$, 
while the loop effects of the EW interactions 
are expected to be at most ${\cal O}(\mbox{few \%})$.
In fact, the sensitivity of $M_{t'}$ and $M_{b'}$ on 
$y_\Lambda^2$ is 20\%--10\% for $\Lambda^{(4)}=2\mbox{--}10$ TeV.
On the other hand, the mass $M_{H_1}$ 
of the $H_1$ Higgs boson varies about 20\% 
for $\Lambda^{(4)} = 2\mbox{--}10$ TeV, 
while the sensitivity of the masses of other Higgs bosons is
at most 5\%.
Taking into account these uncertainties,
one can safely ignore the EW one-loop corrections.

Since the two charged Higgs bosons couple to $t$ and $b$ quarks,
their masses are severely constrained by $R_b$. Moreover,
because in our model $M_{H_{1,2}^\pm}$ are determined by $M_{A_{1,2}}$,
it leads to a constraint for $M_{A_{1,2}}$.
The $2\sigma$-bound of $R_b$ yields 
$M_{A_2} \geq 0.70,0.58,0.50$ TeV for $\Lambda^{(4)}=2,5,10$ TeV,
$\Lambda^{(3)}/\Lambda^{(4)}=1.5$, and $M_{A_1} > 0.1$ TeV.
We note that 
the above constraint for $M_{A_2}$ is not very sensitive to
the values of $M_{A_1}$ and $\Lambda^{(3)}/\Lambda^{(4)}$. 

The $S$ and $T$ parameters for a multiple Higgs doublet model
are analyzed in Ref.~\cite{Grimus:2007if}.
In our model, the Higgs contributions are
$S_h=0\mbox{--}0.1$ and $T_h=-0.02\mbox{--}-0.2$
for $\Lambda^{(4)}=2\mbox{--}10$ TeV, 
$\Lambda^{(3)}/\Lambda^{(4)}=1-2$, $0.1 < M_{A_1} < 0.6$ TeV,
and $0.5 < M_{A_2} < 0.8$ TeV. 
Since the Higgs contribution to the $T$-parameter is 
slightly negative, the mass differences of the fermions, 
depending on the values of $T_h$, are allowed.
For example, following the $(S,T)$ analysis a la LEP EWWG~\cite{LEP:2005ema},
we found that 
our model is within 
the 95\% C.L. contour of the $(S,T)$ constraint,
when the fourth family lepton masses are
$M_{\tau'} - M_{\nu'} \sim 150$~GeV.

A noticeable feature of the presence of the
fourth family is that 
because of the extra loop contributions of $t'$ and $b'$,
the lightest CP even Higgs boson production via the gluon fusion
is considerably enhanced.
For example,
for $\Lambda^{(4)}=3$~TeV, $\Lambda^{(3)}/\Lambda^{(4)}=1.5$,
$\tan\beta_{4}=1$, $M_{A_1}=0.50$~TeV, and $M_{A_2}=0.80$~TeV, 
we obtain $M_{t'}=M_{b'}=0.33$~TeV and $M_{H_1}=0.49$~TeV. In this 
case, the enhancement factor of 
$\sigma_{gg \to H_1} \mbox{Br}(H_1 \to ZZ)$ to the SM value is $5.1$,
where the relative $H_1 ZZ$ and $H_1 t\bar{t}$ couplings to the SM values
are $0.86$ and $2.0$, respectively.
Similarly, the CP odd Higgs production via the gluon fusion
should be also enhanced, compared with $gg \to A$ 
in the two Higgs doublet model. 

\section{CKM structure and flavor changing neutral current processes}
\label{5}

We use the same approach to constructing the CKM matrix as in
Ref.~\cite{Hashimoto:2009xi}. The Yukawa interactions take the form
\begin{equation}
  - {\cal L}_Y =
    \sum_{i,j}\bar{\psi}_L^{(i)} Y_D^{ij} d_R^{(j)} \tilde{\Phi}_{b'}
  + \sum_{i,j}\bar{\psi}_L^{(i)} Y_U^{ij} u_R^{(j)} \Phi_{t'}
  + y_t \bar{\psi}_L^{(3)} t_R \Phi_{t} ,
\end{equation}
where
\begin{equation}
  Y_D \equiv \frac{\sqrt{2}}{v_{b'}} M_D, \quad
  Y_U \equiv \frac{\sqrt{2}}{v_{t'}} M_U, 
\end{equation}
and
\begin{equation}
  M_D =
  \left(\begin{array}{cccc}
    m_d & \xi_{12} m_d & \xi_{13} m_d & \xi_{14} m_d \\
    \xi_{21} m_d & m_s & \xi_{23} m_s & \xi_{24} m_s \\
    \xi_{31} m_d & \xi_{32} m_s & m_b & \xi_{34} m_s \\
    \xi_{41} m_d & \xi_{42} m_s & \xi_{43} m_s & m_{b'}
  \end{array}\right) , 
\end{equation}
\begin{equation}
  M_U =
  \left(\begin{array}{cccc}
    m_u & \eta_{12} m_u & \eta_{13} m_u & \eta_{14} m_u \\
    \eta_{21} m_u & m_c & \eta_{23} m_c & \eta_{24} m_c \\
    \eta_{31} m_u & \eta_{32} m_c & \eta_{33} m_c & \eta_{34} m_c \\
    \eta_{41} m_u & \eta_{42} m_c & \eta_{43} m_c & m_{t'}
  \end{array}\right) \, . 
\end{equation}

In accordance with the essence of the composite 
$(2+1)$-Higgs model, we assumed that the top-Higgs is responsible for
the top mass.

The CKM matrix is approximately given by
\begin{equation}
  V_{CKM}^{4 \times 4} \simeq
  \left(\begin{array}{cccc}
    1-\frac{|\xi_{12}|^2}{2}\left(\frac{m_d}{m_s}\right)^2 &
    \xi_{12} \frac{m_d}{m_s} & \xi_{13} \frac{m_d}{m_b} &
   -(\eta_{14} - \eta_{12}\eta_{24})\frac{m_u}{m_{t'}}
   + \xi_{14} \frac{m_d}{m_{b'}} \\
   -\xi_{12}^* \frac{m_d}{m_s} &
    1-\frac{|\xi_{12}|^2}{2}\left(\frac{m_d}{m_s}\right)^2 &
    \xi_{23} \frac{m_s}{m_b} - \eta_{23} \frac{m_c}{m_t} &
   -\eta_{24} \frac{m_c}{m_{t'}} \\
   -\eta_{23}^* \frac{m_c}{m_t} \cdot \xi_{12}^* \frac{m_d}{m_s}
   -(\xi_{13}^* - \xi_{12}^* \xi_{23}^*)\frac{m_d}{m_b} &
   -\xi_{23}^* \frac{m_s}{m_{b}} + \eta_{23}^* \frac{m_c}{m_t}&
    1 & -\eta_{34} \frac{m_c}{m_{t'}} \\
   -\eta_{24}^* \frac{m_c}{m_{t'}} \cdot \xi_{12}^* \frac{m_d}{m_{s}} &
    \eta_{24}^* \frac{m_c}{m_{t'}} & \eta_{34}^* \frac{m_c}{m_{t'}} & 1
  \end{array}\right) \, .
\end{equation}
Notice that $m_d/m_s \sim 0.1 = {\cal O}(V_{us})$,
$m_s/m_b \sim 0.01 = {\cal O}(V_{cb})$, $m_c/m_t \sim 0.01$, and
$m_d/m_b \sim 0.001 = {\cal O}(V_{ub})$.
Thus we can reproduce the CKM structure by
taking $\xi_{ij} = {\cal O}(1)$.
Since the mixing between the fourth family and the others
is suppressed,
$|V_{t'd}| \sim |V_{us}| m_c/m_{t'} \sim {\cal O}(10^{-3})$ and
$|V_{t's}| \sim |V_{t'b}| \sim m_c/m_{t'} \sim {\cal O}(10^{-2})$,
the contribution of the flavor-changing-neutral-current
(FCNC) processes with the fourth family quarks in the $B$-system 
is negligible: $m_{t'}^2 |V_{t'd}^*V_{t'b}|^2 \sim |V_{us}|^2 m_c^4/m_{t'}^2$
for $B_d$ and $m_{t'}^2 |V_{t's}^*V_{t'b}|^2 \sim m_c^4/m_{t'}^2$
for $B_s$.
Similarly, $b \to s \gamma$ and $Z \to \bar{b}b$ via the $t'$-loop
are also suppressed (for a related discussion, see
Ref.~\cite{Hashimoto:2009xi}).
As to the contribution of a box diagram with $t'$ in the $\Delta S = 2$ processes 
in the K system, it is very small due to 
$m^{2}_{t'} |V_{t'd}^* V_{t's}|^2 \sim m^{2}_c |V_{us}|^2 m^{2}_c/m^{2}_{t'}$. 
Note also that 
the contributions of the charged Higgs bosons are negligible,
because their  masses are relatively heavy and the mixing angles are small.

On the other hand, a new tree FCNC 
term appears in the up-quark sector,
so that the $D^0$--$\bar{D}^0$ mixing is potentially dangerous.
Let us estimate this effect. 
In the basis of the fermion mass eigenstates
$U_{L,R}$ and $D_{L,R}$, corresponding 
to the left and right-handed up-type quarks and 
the down-type ones, respectively,
there appear the tree FCNC and 
flavor-changing-charged-current (FCCC) 
terms in the Higgs sector:
\begin{eqnarray}
{\cal L}_{\rm FCNC/FCCC}  &=&
 - \frac{m_t}{v_{t'}} \bar{U}_L \tilde{M} U_R (h_{t'} - iz_{t'})
 + \frac{m_t}{v_{t}} \bar{U}_L \tilde{M} U_R (h_{t} - iz_{t})
 + \mbox{(h.c.)} \nonumber  \\
&&
 + \sqrt{2} \frac{m_t}{v_{t'}} \bar{U}_R \tilde{M}^\dagger
   V_{\rm CKM}^{4 \times 4} D_L \omega_{t'}^+
 - \sqrt{2} \frac{m_t}{v_{t}} \bar{U}_R \tilde{M}^\dagger
   V_{\rm CKM}^{4 \times 4} D_L \omega_{t}^+ + \mbox{(h.c.)} ,
\end{eqnarray}
where the fields $h_{t',t}$, $z_{t',t}$, and $\omega_{t',t}^\pm$
are defined in Eq.~(\ref{comp}). The matrix $\tilde{M}$ is
\begin{equation}
  \tilde{M}_{ij} \equiv ({\cal U}_L)_{3i}^* ({\cal U}_R)_{3j} ,
\end{equation}
where 
\begin{equation}
  {\cal U}_L \simeq
  \left(\begin{array}{cccc}
    1 & \eta_{12} \frac{m_u}{m_c} & \eta_{13} \frac{m_u}{m_t} &
    \eta_{14} \frac{m_u}{m_{t'}} \\
   -\eta_{12}^* \frac{m_u}{m_c} & 1 &
    \eta_{23} \frac{m_c}{m_t} & \eta_{24} \frac{m_c}{m_{t'}} \\
   -(\eta_{13}^* - \eta_{12}^* \eta_{23}^*)\frac{m_u}{m_t} &
   -\eta_{23}^* \frac{m_c}{m_t} &
    1 & \eta_{34} \frac{m_c}{m_{t'}} \\
   -(\eta_{14}^* - \eta_{12}^* \eta_{24}^*)\frac{m_u}{m_{t'}} &
   -\eta_{24}^* \frac{m_c}{m_{t'}} & -\eta_{34}^* \frac{m_c}{m_{t'}} & 1
  \end{array}\right) ,
\end{equation}
and
\begin{equation}
  {\cal U}_R \simeq
  \left(\begin{array}{cccc}
    1 & \eta_{21}^* \frac{m_u}{m_c} & \eta_{31}^* \frac{m_u}{m_t} &
    \eta_{41}^* \frac{m_u}{m_{t'}} \\
   -\eta_{21} \frac{m_u}{m_c} & 1 &
    \eta_{32}^* \frac{m_c}{m_t} & \eta_{42}^* \frac{m_c}{m_{t'}} \\
   -(\eta_{31} - \eta_{21} \eta_{32}) \frac{m_u}{m_t} &
   -\eta_{32} \frac{m_c}{m_t} & 1 & \eta_{43}^* \frac{m_c}{m_{t'}} \\
   -(\eta_{41} - \eta_{21} \eta_{42}) \frac{m_u}{m_{t'}} &
   -\eta_{42} \frac{m_c}{m_{t'}} & -\eta_{43} \frac{m_c}{m_{t'}} & 1
  \end{array}\right) \, 
\end{equation}
are the transformation matrices from the weak basis to the mass eigenstates
one for up-type quarks. The dangerous contributions to
the $D^0$-$\bar{D}^0$ mixing come from
the $u$-$c$-$h_{t',t}$ couplings
and thus they are proportional to
\begin{equation}
  Y_{u-c-h_{t',t}} \simeq \frac{m_t}{v_{t',t}}
  \frac{m_u}{m_t}\frac{m_c}{m_t} .
\end{equation}
Therefore the corresponding contribution to the 
$D^0$-$\bar{D}^0$ mixing parameter $\Delta m_D/m_D$ is of order
\begin{equation}
  \frac{Y_{u-c-h_{t',t}}^2}{M_{H_{2,3}}^2} f_D^2 B_D
  \sim \frac{f_D^2 B_D}{M_{H_{2,3}}^2} \times O(10^{-14}) , 
\end{equation}
where $f_D$ is the $D$ meson decay constant,
$B_D$ denotes the $B$ parameter, and
we ignored the mixing between $h_{t',t}$ 
and $H_{1,2,3}$.
Since the experimental value of the $D^0$-$\bar{D}^0$
mixing
parameter is 
$\Delta m_D/m_D \sim O(10^{-14})$ \cite{pdg}
and $f_D \sim O(\mbox{100MeV})$,
this tree FCNC contribution is negligible for
$M_{H_{2,3}}$ of the order of the EWSB scale.
Due to the same reasons, the tree FCCC 
contribution is also suppressed
in the first and second families.

\section{Conclusion}
\label{6}

The 2 + 1 composite Higgs model
is an offspring of the top quark
condensate one \cite{MTY,Nambu,Bardeen:1989ds} but has 
much richer and more sophisticated dynamics. 
As a result, this allows to describe rather naturally both
the quark mass hierarchy \cite{Hashimoto:2009xi} and EWSB. 
It is quite nontrivial that 
this model passes
the electroweak precision data constraints.
Besides, we can naturally evade the constraint of $Z \to b\bar{b}$,
because the top-Higgs is sufficiently heavy.

It is also noticeable that the model has a clear 
signature: the 2 + 1 structure of the composite Higgs bosons. 
In the heaviest 
doublet, the top-Higgs
$\Phi_t \sim \bar{t}_{R}(t,b)_L$ component dominates. 
As is clearly illustrated by Figs. \ref{ma1-ma2} and \ref{ma2-dep},
the masses of the four resonances in this doublet are
nearly degenerate that reflects a subcritical dynamics of the t quark. 
 
Other phenomenological manifestations of the model are
the following. The gluon-fusion channel with a decay to two $Z$ bosons
should be essentially enhanced. For example,
for the parameter set with
$\Lambda^{(4)}=3$~TeV, $\Lambda^{(3)}/\Lambda^{(4)}=1.5$,
$\tan\beta_{4}=1$, $M_{A_1}=0.50$~TeV, and $M_{A_2}=0.80$~TeV 
(which yields $M_{t'}=M_{b'}=0.33$~TeV and $M_{H_1}=0.49$~TeV),
the enhancement factor of 
$\sigma_{gg \to H_1} \mbox{Br}(H_1 \to ZZ)$ to the SM value is $5.1$.
Similarly, the CP odd Higgs production via the gluon fusion
should be enhanced as well, compared with a three family model.
Also, multiple Higgs bosons may be observed as 
$t\bar{t}$ resonances at the LHC~\cite{Frederix:2007gi}.
Detailed analysis of their LHC signatures
will be performed elsewhere.

\acknowledgments

The research of M.H. was supported by
the Grant-in-Aid for Science Research, Ministry of Education, 
Culture, Sports, Science and Technology, Japan, No. 16081211.
The work of V.A.M. was supported by the Natural Sciences and
Engineering Research Council of Canada.

\appendix

\section{Bubble approximation}
\label{bubble-app}

In the bubble approximation, i.e., 
the $1/N$-leading approximation neglecting the QCD effects,
we can easily obtain the low energy effective theory;
\begin{equation}
  {\cal L} = {\cal L}_f + {\cal L}_g + {\cal L}_s + {\cal L}_y, 
\end{equation}
with
\begin{equation}
 {\cal L}_s = |D_\mu \Phi_{b'}|^2 + |D_\mu \Phi_{t'}|^2
 + |D_\mu \Phi_{t}|^2 - V,
\end{equation}
\begin{equation}
 - {\cal L}_y = 
   y_{q'} \bar{\psi}_L^{(4)} {\cal M}_{\Phi_{t'} \Phi_{b'}} \psi_R^{(4)} 
 + y_t \bar{\psi}_L^{(3)}t_R \Phi_{t} + \mbox{(h.c.)},
\end{equation}
and
\begin{eqnarray}
  V &=& V_2 + V_4, \\
V_2 &=& 
   M_{\Phi_{b'}}^2 (\Phi_{b'}^\dagger \Phi_{b'})
 + M_{\Phi_{t'}}^2 (\Phi_{t'}^\dagger \Phi_{t'})
 + M_{\Phi_{t}}^2 (\Phi_{t}^\dagger \Phi_{t}) \nonumber \\
&&
 + M_{\Phi_{t'}\Phi_{b'}}^2 (\Phi_{t'}^\dagger \Phi_{b'})
 + M_{\Phi_{t'}\Phi_{t}}^2 (\Phi_{t'}^\dagger \Phi_{t})
 + M_{\Phi_{b'}\Phi_{t}}^2 (\Phi_{b'}^\dagger \Phi_{t})
 + \mbox{(h.c.)}, \\
V_4 &=&
   \lambda \tr({\cal M}_{\Phi_{t'} \Phi_{b'}}^\dagger 
               {\cal M}_{\Phi_{t'} \Phi_{b'}})^2
 + \frac{1}{2}\lambda_t \tr({\cal M}_{\Phi_{t}}^\dagger {\cal M}_{\Phi_{t}})^2,
\end{eqnarray}
where we have already renormalized the composite scalar fields
and also defined $2 \times 2$ Higgs fields,
\begin{equation}
  {\cal M}_{\Phi_{t'} \Phi_{b'}} \equiv (\Phi_{t'} \; \tilde{\Phi}_{b'}), 
  \quad
  {\cal M}_{\Phi_{t}} \equiv (\Phi_{t} \; \tilde{\Phi}_{t}), 
\end{equation}
and the right-handed doublet
\begin{equation}
  \psi_R^{(4)} \equiv
  \left(\begin{array}{c} t'_R \\ b'_R \end{array}\right) \, .
\end{equation}
Note that 
\begin{equation}
    \tr({\cal M}_{\Phi_{t'} \Phi_{b'}}^\dagger {\cal M}_{\Phi_{t'} \Phi_{b'}})^2
   = (\Phi_{b'}^\dagger \Phi_{b'})^2 + (\Phi_{t'}^\dagger \Phi_{t'})^2
   +2 (\Phi_{b'}^\dagger \Phi_{b'})(\Phi_{t'}^\dagger \Phi_{t'})
   -2 |\Phi_{b'}^\dagger \Phi_{t'}|^2 \, . 
\end{equation}
The renormalized quantities are given by
\begin{eqnarray}
  y_{q'}(\mu) & \equiv & y_{t'}(\mu) = y_{b'}(\mu) =
  \left(\frac{N}{16\pi^2}\ln \frac{(\Lambda^{(4)})^2}{\mu^2}\right)^{-1/2},\\
  y_{t}(\mu) &=&
  \left(\frac{N}{16\pi^2}\ln \frac{(\Lambda^{(3)})^2}{\mu^2}\right)^{-1/2},\\
  M_{\Phi_{t'}}^2(\mu) &=& 
  y_{t'}^2 \left[\,M_{\Phi_{t'0}}^2
                - \frac{N}{8\pi^2}((\Lambda^{(4)})^2 - \mu^2)\,\right],\\
  M_{\Phi_{b'}}^2(\mu) &=& 
  y_{b'}^2 \left[\,M_{\Phi_{b'0}}^2
                - \frac{N}{8\pi^2}((\Lambda^{(4)})^2 - \mu^2)\,\right],\\
  M_{\Phi_{t}}^2(\mu) &=& 
  y_{t}^2 \left[\,M_{\Phi_{t0}}^2
                - \frac{N}{8\pi^2}((\Lambda^{(3)})^2 - \mu^2)\,\right],\\
  M_{\Phi_{t'}\Phi_{b'}}^2 &=& y_{t'} y_{b'} M_{\Phi_{t'0}\Phi_{b'0}}^2, \\
  M_{\Phi_{t'}\Phi_{t}}^2 &=& y_{t'} y_{t} M_{\Phi_{t'0}\Phi_{t0}}^2, \\
  M_{\Phi_{b'}\Phi_{t}}^2 &=& y_{t} y_{b'} M_{\Phi_{b'0}\Phi_{t0}}^2, \\
  \lambda &=& y_{q'}^2, \\
  \lambda_t &=& y_{t}^2 \, .  
\end{eqnarray}

The part of $V_4$ has the global symmetry,
\begin{equation}
  SU(2)_{L4} \times SU(2)_{R4} \times SU(2)_{Lt} \times SU(2)_{Rt}
  \times U(1)_A,
\end{equation}
where the transformation property is
\begin{equation}
  {\cal M}_{\Phi_{t'} \Phi_{b'}} \to 
  g_{L4} {\cal M}_{\Phi_{t'} \Phi_{b'}} g_{R4}^\dagger, 
  \quad
  {\cal M}_{\Phi_{t}} \to g_{Lt} {\cal M}_{\Phi_{t}} g_{Rt}^\dagger,
\end{equation}
with $g_{L4} \in SU(2)_{L4}$, $g_{R4} \in SU(2)_{R4}$,
$g_{Lt} \in SU(2)_{Lt}$ and $g_{Rt} \in SU(2)_{Rt}$.
The hypercharge $U(1)_Y$ is included in the $U(1)$ parts of 
$SU(2)_{R4}$ and $SU(2)_{Rt}$,
\begin{equation}
  {\cal M}_{\Phi_{t'} \Phi_{b'}} \to
  {\cal M}_{\Phi_{t'} \Phi_{b'}}e^{-i\theta_Y \frac{\tau_3}{2}}, \quad
  {\cal M}_{\Phi_{t}} \to {\cal M}_{\Phi_{t}} e^{-i\theta_Y\frac{\tau_3}{2}} , 
\end{equation}
and the $U(1)_A$ corresponds to
\begin{equation}
  {\cal M}_{\Phi_{t'} \Phi_{b'}} \to
  {\cal M}_{\Phi_{t'} \Phi_{b'}}e^{-2i\theta_A}, \quad
  {\cal M}_{\Phi_{t}} \to {\cal M}_{\Phi_{t}} e^{-2i\theta_A\tau_3} \, .
\end{equation}

Beyond the bubble approximation,
another $SU(2)_{R4}$ symmetric coupling, $\tilde{\lambda} 
[\tr({\cal M}_{\Phi_{t'} \Phi_{b'}}^\dagger {\cal M}_{\Phi_{t'} \Phi_{b'}})]^2$, 
is generated at low energy. 
This is the reason why we consider more general expressions 
in Sec.~\ref{L-eff}.

Since there is no bottom Yukawa coupling,
the $SU(2)_{Rt}$ symmetry is explicitly broken down
in the Yukawa sector.
Moreover, the Higgs mass mixing terms $V_2$ respect only 
the $SU(2)_L \times U(1)_Y$ gauge symmetry.
Thus the Higgs mass spectrum does not have 
the $SU(2)_{R4}$ and $SU(2)_{Rt}$ symmetries
in general.

\section{General three Higgs doublet renormalizable model}
\label{3higgs}

Let us consider a potential of a general three Higgs doublet 
renormalizable model
with the scalars $\phi_{1,2,3}$.

The most general potential is 
\begin{equation}
  V = V_2 + V_4,
\end{equation}
where the mass terms,
\begin{equation}
  V_2 \equiv m_1^2 \phi_1^\dagger \phi_1
  + m_2^2 \phi_2^\dagger \phi_2 + m_3^2 \phi_3^\dagger \phi_3
  + [m_{12}^2 \phi_1^\dagger \phi_2 + \mbox{(h.c)}]
  + [m_{23}^2 \phi_2^\dagger \phi_3 + \mbox{(h.c)}]
  + [m_{31}^2 \phi_3^\dagger \phi_1 + \mbox{(h.c)}],
\label{V2}
\end{equation}
and the quartic couplings,
\begin{eqnarray}
  V_4 &\equiv& \phantom{+}
    \lambda_{1111} (\phi_1^\dagger \phi_1)^2
  + \lambda_{2222} (\phi_2^\dagger \phi_2)^2
  + \lambda_{3333} (\phi_3^\dagger \phi_3)^2 \nonumber \\
&& 
  + \lambda_{1122} (\phi_1^\dagger \phi_1)(\phi_2^\dagger \phi_2)
  + \lambda_{2233} (\phi_2^\dagger \phi_2)(\phi_3^\dagger \phi_3)
  + \lambda_{3311} (\phi_3^\dagger \phi_3)(\phi_1^\dagger \phi_1)
    \nonumber \\
&&
  + \lambda_{1221} |\phi_1^\dagger \phi_2|^2
  + \lambda_{2332} |\phi_2^\dagger \phi_3|^2
  + \lambda_{3113} |\phi_3^\dagger \phi_1|^2
    \nonumber \\
&&
  + [\lambda_{1212} (\phi_1^\dagger \phi_2)^2 + \mbox{(h.c)}]
  + [\lambda_{2323} (\phi_2^\dagger \phi_3)^2 + \mbox{(h.c)}]
  + [\lambda_{3131} (\phi_3^\dagger \phi_1)^2 + \mbox{(h.c)}]
    \nonumber \\
&& 
  + [\lambda_{1112} (\phi_1^\dagger \phi_1)(\phi_1^\dagger \phi_2)
     + \mbox{(h.c)}]
  + [\lambda_{1113} (\phi_1^\dagger \phi_1)(\phi_1^\dagger \phi_3)
     + \mbox{(h.c)}]
    \nonumber \\
&&
  + [\lambda_{2221} (\phi_2^\dagger \phi_2)(\phi_2^\dagger \phi_1)
     + \mbox{(h.c)}]
  + [\lambda_{2223} (\phi_2^\dagger \phi_2)(\phi_2^\dagger \phi_3)
     + \mbox{(h.c)}]
    \nonumber \\
&&
  + [\lambda_{3331} (\phi_3^\dagger \phi_3)(\phi_3^\dagger \phi_1)
     + \mbox{(h.c)}]
  + [\lambda_{3332} (\phi_3^\dagger \phi_3)(\phi_3^\dagger \phi_2)
     + \mbox{(h.c)}]
    \nonumber \\
&&
  + [\lambda_{1123} (\phi_1^\dagger \phi_1)(\phi_2^\dagger \phi_3)
     + \mbox{(h.c)}]
  + [\lambda_{1213} (\phi_1^\dagger \phi_2)(\phi_1^\dagger \phi_3)
     + \mbox{(h.c)}]
  + [\lambda_{1231} (\phi_1^\dagger \phi_2)(\phi_3^\dagger \phi_1)
     + \mbox{(h.c)}]
    \nonumber \\
&&
  + [\lambda_{2213} (\phi_2^\dagger \phi_2)(\phi_1^\dagger \phi_3)
     + \mbox{(h.c)}]
  + [\lambda_{2123} (\phi_2^\dagger \phi_1)(\phi_2^\dagger \phi_3)
     + \mbox{(h.c)}]
  + [\lambda_{2132} (\phi_2^\dagger \phi_1)(\phi_3^\dagger \phi_2)
     + \mbox{(h.c)}]
    \nonumber \\
&&
  + [\lambda_{3312} (\phi_3^\dagger \phi_3)(\phi_1^\dagger \phi_2)
     + \mbox{(h.c)}]
  + [\lambda_{3132} (\phi_3^\dagger \phi_1)(\phi_3^\dagger \phi_2)
     + \mbox{(h.c)}]
  + [\lambda_{3123} (\phi_3^\dagger \phi_1)(\phi_2^\dagger \phi_3)
     + \mbox{(h.c)}] \, .
\label{V4}
\end{eqnarray}
For the mass terms, the number of the real parameters is 
\begin{equation}
  N_M = N_H^2 - (N_H-1)\,  , 
\end{equation}
where $N_H$ denotes the number of the Higgs doublets and 
we used the rephasing degrees of freedom of the Higgs fields.
For the quartic couplings, the number of the real parameters is
\begin{equation}
  N_Q = \frac{1}{2} N_H^2(N_H^2+1) \, . 
\end{equation}
In particular, the formula for the two and three Higgs doublets 
yields 10 and 45, which agree with expressions (\ref{V2}) and 
(\ref{V4}) above.

The RGE's for the multi-Higgs models are discussed in 
Ref.~\cite{Komatsu:1981xh}.
The $S$ and $T$ parameters for the multi Higgs doublet model
are analyzed in Ref.~\cite{Grimus:2007if}.

\section{Analytic mass formulas 
for $M_{A_{1,2}}$ and $M_{H_{1,2}^\pm}$}
\label{analy-mass}

Let us define the mixing angles of the CP odd and charged Higgs fields,
\begin{equation}
  \left(\begin{array}{c} z_{b'} \\ z_{t'} \\ z_t \end{array}\right)
  = O^A
  \left(\begin{array}{c} \pi_z \\ A_1 \\ A_2 \end{array}\right), \quad
  \left(\begin{array}{c} w_{b'}^\pm \\ w_{t'}^\pm \\ w_t^\pm \end{array}\right)
  = U^{H^\pm}
  \left(\begin{array}{c} \pi_w^\pm \\ H_1^\pm \\ H_2^\pm \end{array}\right),
\end{equation}
where $\pi_z$ and $\pi_w^\pm$ denote the would-be NG bosons
eaten by the weak bosons.
The mass eigenvalues of the corresponding Higgs mass matrices
are $M_{A_{1,2}}$ and $M_{H_{1,2}^\pm}$.

Eliminating $M_{\Phi_{b'}}^2$, $M_{\Phi_{t'}}^2$ and $M_{\Phi_{t}}^2$
by using the stationary conditions, we obtain the mass eigenvalues of 
the CP odd and charged Higgs bosons:
\begin{eqnarray}
  M_{A_{1,2}}^2 &=&
  \frac{1}{2}\left(\,\frac{-M_{\Phi_{t'} \Phi_{b'}}^2}{s_{\beta_4}c_{\beta_4}}
  + \frac{-M_{\Phi_{t'} \Phi_{t}}^2}{s_{\beta_4}s_{\beta_{34}}c_{\beta_{34}}}
   (s_{\beta_{34}}^2 + s_{\beta_4}^2 c_{\beta_{34}}^2)
  + \frac{-M_{\Phi_{b'} \Phi_{t}}^2}{c_{\beta_4}s_{\beta_{34}}c_{\beta_{34}}}
   (s_{\beta_{34}}^2 + c_{\beta_4}^2 c_{\beta_{34}}^2)
  \mp M_{A_2-A_1}^2\,\right), \\
  (M_{A_2-A_1}^2)^2 &\equiv&
    \left(\,\frac{-M_{\Phi_{t'} \Phi_{b'}}^2}{s_{\beta_4}c_{\beta_4}}
  + \frac{-M_{\Phi_{t'} \Phi_{t}}^2}{s_{\beta_4}s_{\beta_{34}}c_{\beta_{34}}}
   (s_{\beta_{34}}^2 c_{\beta_4}^2 - s_{\beta_{4}}^2)
  + \frac{-M_{\Phi_{b'} \Phi_{t}}^2}{c_{\beta_4}s_{\beta_{34}}c_{\beta_{34}}}
   (s_{\beta_{34}}^2 s_{\beta_4}^2 - c_{\beta_{4}}^2)\,\right)^2
   \nonumber \\ 
&&
  + \frac{4}{c_{\beta_{34}}^2}
    \left(M_{\Phi_{t'} \Phi_{t}}^2 c_{\beta_4}
         -M_{\Phi_{b'} \Phi_{t}}^2 s_{\beta_4}\right)^2,
\end{eqnarray}
and
\begin{align}
&  M_{H_{1,2}^\pm}^2 =
  \frac{1}{2}\left(\,\frac{-M_{\Phi_{t'} \Phi_{b'}}^2}{s_{\beta_4}c_{\beta_4}}
  + \frac{-M_{\Phi_{t'} \Phi_{t}}^2}{s_{\beta_4}s_{\beta_{34}}c_{\beta_{34}}}
   (s_{\beta_{34}}^2 + s_{\beta_4}^2 c_{\beta_{34}}^2)
  + \frac{-M_{\Phi_{b'} \Phi_{t}}^2}{c_{\beta_4}s_{\beta_{34}}c_{\beta_{34}}}
   (s_{\beta_{34}}^2 + c_{\beta_4}^2 c_{\beta_{34}}^2)
  - \frac{1}{2}\lambda_4 v^2 c_{\beta_{34}}^2
  \mp M_{H_2^\pm-H_1^\pm}^2\,\right), \\
&  (M_{H_2^\pm-H_1^\pm}^2)^2 \equiv
    \left(\,\frac{-M_{\Phi_{t'} \Phi_{b'}}^2}{s_{\beta_4}c_{\beta_4}}
  + \frac{-M_{\Phi_{t'} \Phi_{t}}^2}{s_{\beta_4}s_{\beta_{34}}c_{\beta_{34}}}
   (s_{\beta_{34}}^2 c_{\beta_4}^2 - s_{\beta_{4}}^2)
  + \frac{-M_{\Phi_{b'} \Phi_{t}}^2}{c_{\beta_4}s_{\beta_{34}}c_{\beta_{34}}}
   (s_{\beta_{34}}^2 s_{\beta_4}^2 - c_{\beta_{4}}^2)
  - \frac{1}{2}\lambda_4 v^2 c_{\beta_{34}}^2\,\right)^2
   \nonumber \\ 
& \hspace*{2.5cm}
  + \frac{4}{c_{\beta_{34}}^2}
    \left(M_{\Phi_{t'} \Phi_{t}}^2 c_{\beta_4}
         -M_{\Phi_{b'} \Phi_{t}}^2 s_{\beta_4}\right)^2.
\end{align}
Although in principle the analytic 
formulas for the mass eigenvalues of 
the CP even Higgs  can be derived, they are too complicated and, therefore, 
not very useful.

For $M_{\Phi_{b'} \Phi_{t}}^2 = 0$, 
the upper bound of $M_{A_1}$ for a given $M_{A_2}$ is obtained as
\begin{equation}
  \frac{M_{A_1}^2}{M_{A_2}^2} <
  1 + 2 \cot^2 \beta_4 \sin^2 \beta_{34}
    - 2\cot\beta_4 \sin\beta_{34}\sqrt{1+\cot^2 \beta_4 \sin^2\beta_{34}}
  \leq 1 , 
\end{equation}
where the equality on the right hand side satisfies only when
$\cot\beta_4 \sin\beta_{34} = 0$
(by definition, $M_{A_1} \leq M_{A_2}$).

The mixing matrices are defined by
\begin{equation}
  O^A \equiv (\mathbf{n}_v\;\mathbf{e}_1^A\;\mathbf{e}_2^A), \quad
  U^{H^\pm} \equiv (\mathbf{n}_v\;\mathbf{e}_1^{H^\pm}\;\mathbf{e}_2^{H^\pm}),
\end{equation}
with 
\begin{eqnarray}
  \mathbf{n}_v &\equiv&
  \left(\frac{v_{b'}}{v}\; \frac{v_{t'}}{v}\; \frac{v_t}{v}\right)^T, \\[3mm] 
&=&
  \left(\begin{array}{c}
    \cos\beta_4 \cos\beta_{34} \\ \sin\beta_4 \cos\beta_{34} \\
    \sin\beta_{34}
  \end{array}\right) ,
\end{eqnarray}
The analytic formulas for the eigenvectors are
\begin{eqnarray}
  \mathbf{e}_1^X &=& \cos\eta_X \mathbf{e}_1 - \sin\eta_X \mathbf{e}_2, \\  
  \mathbf{e}_2^X &=& \sin\eta_X \mathbf{e}_1 + \cos\eta_X \mathbf{e}_2, 
\end{eqnarray}
where $X=A,H^\pm$, 
\begin{equation}
  \mathbf{e}_1 \equiv \left(\begin{array}{c}
   -\sin\beta_4 \\ \cos\beta_4 \\ 0
  \end{array}\right), \quad
  \mathbf{e}_2 \equiv \left(\begin{array}{c}
   -\cos\beta_4 \sin\beta_{34} \\ -\sin\beta_4 \sin\beta_{34} \\
    \cos\beta_{34}
  \end{array}\right) ,
\end{equation}
\begin{equation}
  \tan\eta_X \equiv \frac{M_{X_1}^2 - \kappa}{\rho}, 
\end{equation}
and
\begin{eqnarray}
  \rho &\equiv& c_{\beta_{34}}^{-1}(-M_{\Phi_{b'} \Phi_{t}}^2 s_{\beta_4}
                               +M_{\Phi_{t'} \Phi_{t}}^2 c_{\beta_4}),\\
  \kappa &\equiv& c_{\beta_{34}}^{-1}
  \left(-M_{\Phi_{b'} \Phi_{t}}^2\frac{c_{\beta_4}}{s_{\beta_{34}}}
        -M_{\Phi_{t'} \Phi_{t}}^2\frac{s_{\beta_4}}{s_{\beta_{34}}}\right)\,.
\end{eqnarray}

The approximate expressions for the mixing matrices are:
\begin{equation}
  O^A \simeq 
  \left(\begin{array}{ccc}
   \frac{1}{\sqrt{2}}\left(1-\frac{\tan^2\beta_{34}}{2}\right) &
  -\frac{1}{\sqrt{2}}\left(1+\frac{\tan^2\beta_{34}}{2}\right) & 
  \frac{M_{A_1}^2}{\sqrt{2}M_{A_2}^2} \tan\beta_{34} \\
  \frac{1}{\sqrt{2}}\left(1-\frac{\tan^2\beta_{34}}{2}\right) &
  \frac{1}{\sqrt{2}}\left(1-\frac{3\tan^2\beta_{34}}{2}\right) & 
  -\sqrt{2}\left(1+\frac{M_{A_1}^2}{2M_{A_2}^2}\right)\tan\beta_{34} \\
  \tan\beta_{34} &
  \left(1+\frac{M_{A_1}^2}{M_{A_2}^2}\right)\tan\beta_{34} & 
  1 - \tan^2\beta_{34}
  \end{array}\right),
\end{equation}
and
\begin{equation}
  U^{H^\pm} \simeq 
  \left(\begin{array}{ccc}
   \frac{1}{\sqrt{2}}\left(1-\frac{\tan^2\beta_{34}}{2}\right) &
  -\frac{1}{\sqrt{2}}\left(1+\frac{\tan^2\beta_{34}}{2}\right) & 
  \frac{M_{H_1^\pm}^2}{\sqrt{2}M_{H_2^\pm}^2} \tan\beta_{34} \\
  \frac{1}{\sqrt{2}}\left(1-\frac{\tan^2\beta_{34}}{2}\right) &
  \frac{1}{\sqrt{2}}\left(1-\frac{3\tan^2\beta_{34}}{2}\right) & 
  -\sqrt{2}\left(1+\frac{M_{H_1^\pm}^2}{2M_{H_2^\pm}^2}\right)\tan\beta_{34} \\
  \tan\beta_{34} &
  \left(1+\frac{M_{H_1^\pm}^2}{M_{H_2^\pm}^2}\right)\tan\beta_{34} & 
  1 - \tan^2\beta_{34}
  \end{array}\right),
\end{equation}
up to ${\cal O}(\tan^2\beta_{34})$,
${\cal O}(\tan\beta_{34} M_{A_1}^2/M_{A_2}^2)$ and 
${\cal O}(\tan\beta_{34} M_{H_1^\pm}^2/M_{H_2^\pm}^2)$.

\end{document}